# Oscillatory deviations from Matthiessen's rule due to interacting dislocations

Chu-Liang Fu[1] and Mingda Li[2*]

[1]*School of Mathematics and Statistics, Yunnan University, Kunming 650500, China*
[2]*Department of Mechanical Engineering, MIT, Cambridge, MA 02139, USA*

We theoretically examine the validity of Matthiessen's rule caused by strong dislocation-dislocation interaction using a fully quantized dislocation field, where its degree of deviation is quantified at arbitrary electron energy, dislocation-electron and dislocation-dislocation distances and interaction strengths. Contrary to intuition, we show that the electron relaxation rate deviates from the Matthiessen's rule in an oscillatory way as a function of inter-dislocation distance, instead of monotonically. This study could serve as a computational tool to investigate the electronic behavior of a highly-dislocated system at a full quantum field theoretical level.

**I. Introduction**

Matthiessen's rule is an empirical relation stating that the total electron relaxation rate in metals and semiconductors can be written as the sum of the scattering rate from each scattering channel, if the interaction between different scattering channels is negligible[1,2]. For instance, when neglecting phonon-impurity and phonon-dislocation interactions, the total electron relaxation rate $1/\tau_{tot}$ accordingly to Matthiessen's rule can be written as $1/\tau_{tot} = 1/\tau_{ph} + 1/\tau_{imp} + 1/\tau_{dis}$, where $1/\tau_{ph}$, $1/\tau_{imp}$ and $1/\tau_{dis}$ denotes the electron relaxation rate from electron-phonon, electron-impurity and electron-dislocation interactions, respectively. The Matthiessen's rule greatly facilitates the calculation of electron mobility, yet as an empirical rule, there are also a few situations for which deviations from Matthiessen's rule could occur, including alloys, rapidly quenched crystals, or heavily plastic deformed materials, etc., where electrons are being scattered strongly by defects, such as strongly-interacting dislocations[3].

On the other hand, despite the experimental awareness of the deviation from Matthiessen's rule, a quantitative evaluation of its applicability caused by strongly-interacting dislocations is still lacking. The reason can be traced back to the lack of a proper methodology taking into account the electron-dislocation scattering and the dislocation-dislocation scattering from an equal footing: in most of the electron-dislocation scattering studies, the first-order perturbation theory is adopted, leading to the proportionality between the electron-dislocation relaxation rate $1/\tau_{dis}$ and the total number of non-interacting dislocations $N_{dis}$, i.e. $1/\tau_{dis} \propto N_{dis}$ [4-6], where the dislocation-dislocation interaction is considered as a higher order term and is simply neglected.

The linearity relation between $1/\tau_{dis}$ and $N_{dis}$ is not limited to dislocation-electron scattering, but also appears in considering the dislocation-phonon scattering [7,8] for the same reason caused by the corresponding perturbative approach. However, a recent non-perturbative functional integral study[9] has shown a breakdown of this perturbative approach due to the strong, long-range interaction between a phonon and a dislocation. The non-perturbative nature for the dislocation-phonon interaction provides an additional incentive to study the validity of Matthiessen's rule considering the influence of interacting dislocations.

Very recently, a fully quantized dislocation field theory has been developed, which is capable of treating the dislocation strain field scattering and the dynamic fluttering scattering on an equal footing using a field theoretical approach, and leads to a new quasiparticle named the "dislon" which can be considered as a quantized lattice displacement field near a dislocation line[10,11]. The dislon theory provides a proper theoretical tool going beyond the lowest order perturbation theory, hence allows us to formulate the electron-dislocation interaction and dislocation-dislocation interaction within a unified framework at a full quantum field theoretical level.

In the study presented below, we examine the Matthiessen's rule for the independent dislocation scattering(IDS), that $1/\tau_{dis} = N_{dis}/\tau_{dis}^1$, where $\tau_{dis}^1$ is the relaxation rate for a single dislocation, by generalizing the original dislon theory[11] to the case of incorporating the dislocation-dislocation interaction. Using a Green's function's approach to solve the propagator of the interacting dislon, we are able to provide quantitative answers to a number of questions regarding the validity, applicability and breakdown conditions of Matthiessen's rule, including the following situations:

1) When an electron is approaching a dislocation core, how would the Matthiessen's rule for the IDS change as a function of electron-dislocation distance?

2) Does the deviation from the Matthiessen's rule for the IDS increase monotonically as the dislocation density increases?

3) Under which condition does the electron relaxation rate significantly deviate from the Matthiessen's rule for the IDS?

By answering these questions, this study not only clarifies the applicability of Matthiessen's rule for IDS at a quantitative level, but also sheds light on further computations of the electronic transport properties in a highly-dislocated crystal.

## II. Quantized theory of a non-interacting dislocation: dislon

To begin with, we review the quantized dislocation theory for a single dislocation line, which is proposed in [11] and is the basis prior to consider dislocation-dislocation interaction. For a single dislocation line in an isotropic medium, with the dislocation core located at $(x_0, y_0) = (0,0)$ and extending along the $z$-direction, denoting the displacement of the dislocation line at position $z$ as $Q(z)$, then we write the mode expansion [12] as

$$Q(z) = \sum_\kappa Q_\kappa e^{i\kappa z} \quad (1)$$

where $\kappa$ is the wavenumber along the dislocation line direction. Defining $\mathbf{u}(\vec{R})$ as the lattice displacement at the spatial lattice point $\vec{R}$, i.e. the difference of the atomic position between a dislocated lattice and a perfect lattice, the $i^{th}$ component of $\mathbf{u}(\vec{R})$ can be expanded as[11,12]

$$\begin{aligned} u_i(\vec{R} \equiv (x,y,z)) &= \sum_\kappa f_i(x,y;\kappa) e^{i\kappa z} Q_\kappa \\ &= \frac{1}{L^2} \sum_{\vec{k}} B_i(\vec{k} \equiv (\vec{s};\kappa)) e^{i\vec{k}\cdot\vec{R}} Q_\kappa \end{aligned} \quad (2)$$

where $L$ is the sample length, and the expansion coefficients $B_i(\vec{k})$ can be written as

$$B_i(\vec{k}) = +\frac{i}{k^2} \left( \begin{array}{c} n_i(\vec{b}\cdot\vec{k}) + b_i(\vec{n}\cdot\vec{k}) \\ -\frac{1}{(1-\nu)} \frac{k_i(\vec{n}\cdot\vec{k})(\vec{b}\cdot\vec{k})}{k^2} \end{array} \right) \quad (3)$$

in which $\vec{b}$ is the Burgers vector, $\vec{n}$ is slip-plane normal direction, and $\nu$ is the Poisson ratio. Using this expansion, the classical 3D kinetic energy and potential energy of this dislocation line can further be written as an effective 1D Hamiltonian after integrating the planar directions as [12,13]

$$H = \frac{L}{2} \sum_\kappa \frac{P_\kappa P_\kappa^*}{m(\kappa)} + \frac{L}{2} \sum_\kappa \kappa^2 K(\kappa) Q_\kappa Q_\kappa^* \quad (4)$$

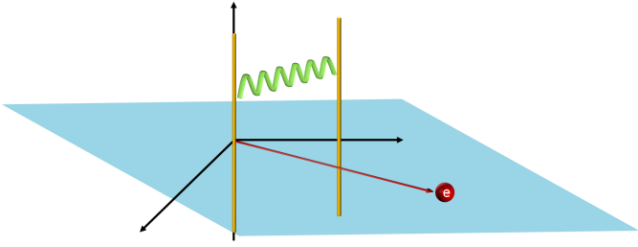

**Fig. 1.** (Color Online). The schematics of the interaction between an electron (red sphere) and two interacting dislocation lines (a and b, golden lines), where the dislocation-dislocation interaction is denoted as green wavy lines, and the alternating blue and red fields denote the schematic total stress field of the two dislocations.

where $P_\kappa$ is the canonical momentum conjugate to $Q_\kappa$, and the mass-like coefficient $m(\kappa)$ and spring-constant-like coefficient $K(\kappa)$ can be written in closed form for an edge dislocation or a screw dislocation, respectively[11]. For an edge dislocation along the $z$-direction (subscript $E$), we have

$$m_E(\kappa) = \frac{\rho b^2}{4\pi} \left[ \begin{array}{l} \ln\left(1+\frac{k_D^2}{\kappa^2}\right) - \frac{k_D^2}{k_D^2+\kappa^2} + \frac{4\nu-3}{8(1-\nu)^2} \times \\ \left( \ln\left(1+\frac{k_D^2}{\kappa^2}\right) - \frac{k_D^2(3k_D^2+2\kappa^2)}{2(k_D^2+\kappa^2)^2} \right) \end{array} \right] \quad (5)$$

$$K_E(\kappa) = \frac{\mu b^2}{4\pi} \left[ \frac{1-2\nu}{2(1-\nu)} \ln\frac{k_D^2}{\kappa^2} + 1 - \frac{1}{4(1-\nu)} \frac{\kappa^2}{k_D^2} \right]$$

where $b = |\vec{b}|$ is the length of Burgers vector, $\rho$ is the mass density, $\mu$ is 2nd-Lamé constant (shear modulus) and $k_D$ is the Debye wavenumber cutoff.

As to a screw dislocation (subscript $S$), we have

$$m_S(\kappa) = \frac{\rho b^2}{4\pi} \left[ \begin{array}{l} \frac{k_D^2}{2(\kappa^2+k_D^2)} + \frac{1}{2}\ln\left(1+\frac{k_D^2}{\kappa^2}\right) \\ + \frac{4\nu-3}{4(1-\nu)^2} \frac{k_D^4}{(k_D^2+\kappa^2)^2} \end{array} \right] \quad (6)$$

$$K_S(\kappa) = \frac{\mu b^2}{4\pi} \left[ \frac{1+\nu}{2(1-\nu)}\left(\ln\frac{k_D^2}{\kappa^2} - 1\right) + \frac{1}{1-\nu}\frac{\kappa^2}{k_D^2} \right]$$

A fully quantized dislocation Hamiltonian can be obtained applying from a canonical quantization procedure to Eq. (4), that imposes the canonical quantization condition,

$$[Q_\kappa, P_{\kappa'}] = i\hbar \delta_{\kappa,\kappa'} \quad (7)$$

From these definitions, we further define the dislon annihilation and creation operators $d_\kappa$ and $d_\kappa^+$ as

$$\begin{cases} Q_\kappa = \sqrt{\frac{\hbar}{2Lm(\kappa)\omega(\kappa)}} \left[ d_\kappa + d_{-\kappa}^+ \right] \\ P_\kappa = i\sqrt{\frac{L\hbar m(\kappa)\omega(\kappa)}{2}} \left[ d_\kappa^+ - d_{-\kappa} \right] \end{cases} \quad (8)$$

where $\omega(\kappa) \equiv v_s \kappa \sqrt{K(\kappa)/m(\kappa)}$, in which $v_s \equiv \sqrt{\mu/\rho}$ is the shear velocity. Substituting Eq. (8) back into Eq. (4), the classical dislocation Hamiltonian with kinetic and potential energy can be rewritten as a collection of independent Bosonic excitations, called "dislons"[11]:

$$H_D = \sum_\kappa \omega(\kappa) \left[ d_\kappa^+ d_\kappa + \frac{1}{2} \right] \quad (9)$$

Where the dislon annihilation and creation operators $d_\kappa$ and $d_\kappa^+$ satisfy Bosonic commutation relation $[d_\kappa, d_{\kappa'}^+] = \delta_{\kappa\kappa'}$.

Eq. (9) is the fundamental quantum Hamiltonian of a single dislocation line in an isotropic medium.

## III. Quantized theory of interacting dislons

Now we generalize the single dislocation theory Eqs. (1)-(9) to the case of interacting dislocations. In a weakly dislocated crystal with a dislocation number density ~$10^6/cm^2$, the effect of dislocation-dislocation interaction is expected to be extremely low; the interaction effect is only apparent in either a heavily dislocated crystal, or at grain boundaries which are composed of a well-defined array of dislocations [14,15]. The latter case is not our primary interest in this study since grain boundaries are generally considered as 2D planar defects instead of 1D line defects. Thus we will focus on the former case of interacting dislocations. On one hand, a single dislocation line in an infinitely-large crystal has infinite strain energy, and therefore dislocation pairs are necessary to regulate the divergence [16]. On the other hand, dislocation pairs occur frequently, such as in a partial dislocation slip [17], in a strain relaxation process[18], and in a crack tip [19,20]. Therefore, we will study the interacting dislocation pairs using a pair wise interaction.

The total Hamiltonian for a system consisting of two interacting dislocation lines (namely, *a* and *b*, See in Fig. 1) can be written as:

$$H = H_a + H_b + H_{int} \tag{10}$$

From Eq. (9), the non-interacting Hamiltonians for the two dislocations can be written down directly as:

$$H_a = \sum_\kappa \omega_a(\kappa)\left(a_\kappa^+ a_\kappa + \frac{1}{2}\right)$$
$$H_b = \sum_\kappa \omega_b(\kappa)\left(b_\kappa^+ b_\kappa + \frac{1}{2}\right) \tag{11}$$

where $\omega_a(\kappa)$ and $\omega_b(\kappa)$ are dispersion relations for dislocation *a* and *b*, which are considered as separate quantities. $a_\kappa$ and $b_\kappa$ are the corresponding operators for dislocation *a* and dislocation *b*, respectively.

To construct the interacting Hamiltonian, we notice that the $i^{th}$ component of the individual displacement field $u_i^\tau$ ($\tau = a, b$) caused by each dislocation *a* or *b* with a dislocation core located at $r_\tau \equiv (x_\tau, y_\tau)$ can be written as a direct in-plane translation of Eq. (2) as

$$u_i^\tau(R) = \frac{1}{L^2}\sum_k B_i^\tau(s;\kappa)e^{ik\cdot R}e^{-is\cdot r_\tau}Q_\kappa^\tau, \quad \tau = a,b \tag{12}$$

Now if we adopt the canonical quantization relation from generalizing Eq. (7) for each individual dislocation, so that

$$[Q_{a,\kappa}, P_{b,\kappa'}] = i\delta_{\kappa,\kappa'}\delta_{ab} \tag{13}$$

and defining the creation and annihilation operator same way as Eq. (8), that

$$\begin{cases} Q_{a,\kappa} = Z_{a,\kappa}\left[a_\kappa + a_{-\kappa}^+\right] \\ P_{a,\kappa} = \frac{i\hbar}{2Z_{a,\kappa}}\left[a_\kappa^+ - a_{-\kappa}\right] \end{cases}, \quad \begin{cases} Q_{b,\kappa} = Z_{b,\kappa}\left[b_\kappa + b_{-\kappa}^+\right] \\ P_{b,\kappa} = \frac{i\hbar}{2Z_{b,\kappa}}\left[b_\kappa^+ - b_{-\kappa}\right] \end{cases} \tag{14}$$

where $Z_{\tau,\kappa} = \sqrt{\frac{1}{2Lm_\tau(\kappa)\omega_\tau(\kappa)}}$, then we could obtain the canonical commutation relations for two dislocations,

$$[a_\kappa, b_{\kappa'}] = 0$$
$$[a_\kappa, b_{\kappa'}^+] = \delta_{\kappa,\kappa'}\delta_{ab} \tag{15}$$

The corresponding lattice displacements are obtained by substituting Eq. (14) back into (12), yielding

$$u_i^\tau(x,y,z) = \frac{1}{L^2}\sum_k B_i^\tau(s;\kappa)e^{ik\cdot R}e^{-is\cdot r_\tau} \times \sqrt{\frac{1}{2Lm_\tau(\kappa)\omega_\tau(\kappa)}}\left[\tau_\kappa + \tau_{-\kappa}^+\right], \quad \tau = a,b \tag{16}$$

Now noticing the fact that the total lattice displacement $u_{tot}$ is the vector sum of individual dislocation, $u_{tot} = u_a + u_b$, the dislocation-dislocation interaction arises from the cross term when computing the quadratic kinetic energy and potential energy

$$H_{int} = T_{int} + U_{int}$$
$$= \rho\int\sum_{i=1}^3 \dot{u}_{i,a}\dot{u}_{i,b}dV + \frac{1}{2}\int c_{ijkl}\left(u_{ij,a}u_{kl,b} + u_{ij,b}u_{kl,a}\right)d^3R$$
$$= \frac{\rho}{2L}\sum_{k\equiv(s,\kappa)}\left(B^a(s;\kappa)\cdot B^{b*}(s;\kappa)e^{+is\cdot d}\dot{Q}_\kappa^a\dot{Q}_\kappa^{b*} + b\leftrightarrow a\right) +$$
$$\frac{1}{2L}\sum_{k\equiv(s,\kappa)}\begin{pmatrix}\begin{pmatrix}(\lambda+\mu)(k\cdot B^a(k))(k\cdot B^{b*}(k))\\+\mu k^2 B^a(k)\cdot B^{b*}(k)\end{pmatrix}e^{+is\cdot d}Q_\kappa^a Q_\kappa^{b*}\\+b\leftrightarrow a\end{pmatrix} \tag{17}$$

where $d \equiv r_b - r_a$ is the difference of the dislocation core positions and $c_{ijkl}$ is elastic tensor. Substituting Eq. (14) back to Eq. (17), the interacting Hamiltonian between dislocations *a* and *b* can be rewritten as

$$H_{int} = \sum_\kappa \begin{pmatrix} M_{ab}(\kappa)\left[a_\kappa^+ - a_{-\kappa}\right]\left[b_\kappa - b_{-\kappa}^+\right] \\ + M_{ba}(\kappa)\left[b_\kappa^+ - b_{-\kappa}\right]\left[a_\kappa - a_{-\kappa}^+\right] \\ + K_{ab}(\kappa)\left[a_\kappa + a_{-\kappa}^+\right]\left[b_{-\kappa} + b_\kappa^+\right] \\ + K_{ba}(\kappa)\left[b_\kappa + b_{-\kappa}^+\right]\left[a_{-\kappa} + a_\kappa^+\right] \end{pmatrix} \tag{18}$$

where the anisotropic, distance-dependent inter-dislocation coupling coefficients $M_{ab}(\kappa)$ and $K_{ab}(\kappa)$ can be expressed as:

$$M_{ab}(\kappa) = \frac{\rho}{4L^2} \sqrt{\frac{\omega_a(\kappa)\omega_b(\kappa)}{m_a(\kappa)m_b(\kappa)}} \sum_{s} B^a(s;\kappa) B^{b*}(s;\kappa) e^{+is\cdot d}$$

$$K_{ab}(\kappa) = \frac{1}{L^2} \frac{1}{4\sqrt{m_a(\kappa)\omega_a(\kappa)m_b(\kappa)\omega_b(\kappa)}} \times \sum_{s} \left[ \begin{array}{l} (\lambda+\mu)(\mathbf{k}\cdot B^a(\mathbf{k}))(\mathbf{k}\cdot B^{b*}(\mathbf{k})) \\ +\mu \mathbf{k}^2 B^a(\mathbf{k})\cdot B^{b*}(\mathbf{k}) \end{array} \right] e^{+is\cdot d} \quad (19)$$

The effect of the dislocation-dislocation interaction on the electron transport can thus be understood by solving the interacting Hamiltonian Eqs. (18) and (19).

## IV. Propagator of a dislon with the presence of the inter-dislon interaction

To see how the dislocation-dislocation interaction alters the IDS, we need to compute the dislon propagator under the dislon-dislon interaction. By redefining a new set of operators to simplify the interaction Hamiltonian Eq. (18) through a canonical transformation, that

$$A_\kappa = \frac{1}{\sqrt{2}}(a_\kappa + a^+_{-\kappa}) = A^+_{-\kappa}, B_\kappa = \frac{1}{\sqrt{2}}(a_\kappa - a^+_{-\kappa}) = -B^+_{-\kappa}$$
$$C_\kappa = \frac{1}{\sqrt{2}}(b_\kappa + b^+_{-\kappa}) = C^+_{-\kappa}, D_\kappa = \frac{1}{\sqrt{2}}(b_\kappa - b^+_{-\kappa}) = -D^+_{-\kappa} \quad (20)$$

The non-interacting Hamiltonians $H_a$ and $H_b$ in Eq. (11) and interacting Hamiltonian $H_{int}$ in Eq. (18) can be rewritten as

$$H_a = \frac{1}{2}\sum_\kappa \omega_a(\kappa)\left(A^+_\kappa A_\kappa + B^+_\kappa B_\kappa\right)$$
$$H_b = \frac{1}{2}\sum_\kappa \omega_b(\kappa)\left(C^+_\kappa C_\kappa + D^+_\kappa D_\kappa\right) \quad (21)$$
$$H_{int} = \sum_\kappa M_{ab}(\kappa) B^+_\kappa D_\kappa + K_{ab}(\kappa) A_\kappa C^+_\kappa + h.c.$$

In order to understand the influence of the dislocation-dislocation interaction to individual dislon behavior, we define a set of retarded Green's functions for the two interacting dislocations

$$A_{\kappa q}(t-t') = -i\theta(t-t')\langle [A_\kappa(t), A^+_q(t')]\rangle$$
$$B_{\kappa q}(t-t') = -i\theta(t-t')\langle [B_\kappa(t), A^+_q(t')]\rangle$$
$$C_{\kappa q}(t-t') = -i\theta(t-t')\langle [C_\kappa(t), A^+_q(t')]\rangle \quad (22)$$
$$D_{\kappa q}(t-t') = -i\theta(t-t')\langle [D_\kappa(t), A^+_q(t')]\rangle$$

where $\langle \rangle$ denotes the ensemble average and $\theta(t-t')$ is the time-ordered step function. Now using the Heisenberg equation of motion and the commutation relations, the equations of motion of Green's function in Eq. (22) can be computed as:

$$i\partial_t A_{\kappa q}(t-t') = \omega_a(\kappa) B_{\kappa q}(t-t') + 2MD_{\kappa q}(t-t')$$
$$i\partial_t B_{\kappa q}(t-t') = \delta_{t,t'}\delta_{\kappa,q} + \omega_a(\kappa) G_{\kappa q}(t-t') + 2KC_{\kappa q}(t-t') \quad (23)$$
$$i\partial_t C_{\kappa q}(t-t') = \omega_b(\kappa) D_{\kappa q}(t-t') + 2MB_{\kappa q}(t-t')$$
$$i\partial_t D_{\kappa q}(t-t') = \omega_b(\kappa) C_{\kappa q}(t-t') + 2KG_{\kappa q}(t-t')$$

where we have used a shorthand notation that $M \equiv \text{Re}(M_{ab}(\kappa))$ and $K \equiv \text{Re}(K_{ab}(\kappa))$.

Performing the Fourier transform to the frequency domain, we obtain

$$\omega A_{\kappa q}(\omega) = \omega_a(\kappa) B_{\kappa q}(\omega) + 2MD_{\kappa q}(\omega)$$
$$\omega B_{\kappa q}(\omega) = \frac{1}{2\pi}\delta_{\kappa,q} + \omega_a(\kappa) G_{\kappa q}(\omega) + 2KC_{\kappa q}(\omega) \quad (24)$$
$$\omega C_{\kappa q}(\omega) = \omega_b(\kappa) D_{\kappa q}(\omega) + 2MB_{\kappa q}(\omega)$$
$$\omega D_{\kappa q}(\omega) = \omega_b(\kappa) C_{\kappa q}(\omega) + 2KG_{\kappa q}(\omega)$$

Now Eq. (24) is no more than a set of linear algebraic equations. Solving these equations and making the assumption for two identical dislocations, that $\omega_\kappa = \omega_a = \omega_b$ to simplify the Green's function, we obtain the expression of the full dislon Green's function upon interaction, that

$$A_\kappa(\omega) \equiv A_{\kappa q}(\omega)$$
$$= \frac{1}{2\pi}\delta_{\kappa,q} \frac{\omega_\kappa(\omega^2 - 4MK - \omega_\kappa^2) + 4M\omega_\kappa(M+K)}{(\omega^2 - 4MK - \omega_\kappa^2)^2 - 4\omega_\kappa^2(M+K)^2} \quad (25)$$

## V. Electron-dislon interaction

Now we study the electron self-energy when scattering with two interacting dislocations (Fig. 1). As a three-body problem (1 electron + 2 dislocations), a rigorous solution is not expected. However, using the fully quantized dislon approach to describe dislocations, the effect of dislocation-dislocation interaction on electron transport properties can be studied under the many-body framework. For simplicity, instead of considering the electron relaxation of both dislocations with interaction, we focus on the interaction between the electron and the dislocation *a*, under the presence of the dislocation-dislocation interaction between dislocations *a* and *b*. In this sense, the approximate solution becomes straightforward to obtain since we only need to consider the renormalized dislon propagator, instead of considering both renormalized dislon propagator and electron-dislon vertex. Under 1-loop correction where electron emits a virtual dislon and reabsorbs the same virtual dislon, the self-energy of the electron-dislon scattering can be written based on the following 1-loop Feynman diagram

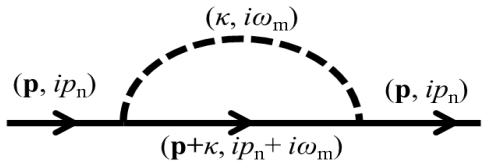

$$\Sigma^{(1)}(\mathbf{r},\mathbf{p},ip_n) = -\frac{1}{\beta L}\sum_{\kappa,\omega_m}|Z_{\mathbf{b},\mathbf{r}}(\kappa)|^2 A_\kappa(i\omega_m) G^{(0)}(\mathbf{p}+\kappa, ip_n+i\omega_m) \quad (26)$$

where **r** is the electron position with momentum *p* related to the dislocation *a*, $A_\kappa(i\omega_m)$ is the full Green's function of dislocation *a* caused by the dislocation-dislocation interaction defined in Eq. (25), $p_n = \frac{(2n+1)\pi}{\beta}$ is the Fermionic Matsubara frequency, $\omega_m = \frac{2m\pi}{\beta}$ is Bosonic Matsubara frequency, and the electron-dislon coupling vertex $Z_{\mathbf{b},\mathbf{r}}(\kappa)$ can be written as[11]:

$$Z_{\mathbf{b},\mathbf{r}}(\kappa) \equiv \frac{Ne^2}{V}\frac{1-2\nu}{1-\nu}\sqrt{\frac{\hbar}{2m(\kappa)\omega(\kappa)}} \times \\ \int s^2 ds \frac{b_x s J_2(rs)\sin 2\phi - 2ib_z \kappa J_1(rs)\sin\phi}{(s^2+\kappa^2+k_{TF}^2)(s^2+\kappa^2)} \quad (27)$$

In the non-interacting case, the full dislon propagator Eq. (25) can be reduced to a free-dislon propagator

$$A^0_{\kappa q}(\omega) = \frac{1}{2\pi}\delta_{\kappa,q}\frac{\omega_\kappa}{\omega^2-\omega_\kappa^2} \quad (28)$$

which can either be seen by directly computing the propagator using the non-interacting dislon Hamiltonian Eq.(9), or we can set the inter-dislocation coupling strength in Eq. (26) to zero ($M_{ab}(\kappa)$=0, $K_{ab}(\kappa)$=0).

With the aid of Cauchy's residual theorem, the electron self-energy Eq. (26) can be further computed as

$$\Sigma^{(1)}(\mathbf{r},\mathbf{p},ip_n) = \frac{1}{L}\sum_\kappa |Z_{\mathbf{b},\mathbf{r}}(\kappa)|^2 S \quad (29)$$

where S can be written as the sum of contributions from 5 different types of 1st-order poles,

$$S = C_1 + C_2 + C_3 + C_4 + C_5 \quad (30)$$

with each coefficient defined as

$$C_1 = \frac{1}{2\pi}\frac{1}{e^{\beta r_1}-1} \times \\ \frac{\omega_\kappa(r_1^2 + 4M^2 - \omega_\kappa^2)}{r_1^4 - (8MK+2\omega_\kappa^2)r_1^2 + (\omega_\kappa^2-4M^2)(\omega_\kappa^2-4K^2)}$$

$$C_{2,3} = \frac{1}{2\pi}\frac{1}{e^{\beta r_{2,3}}-1} \times \\ \frac{1}{ip_n + r_{2,3} - \varepsilon_{\mathbf{p}+\kappa}}\frac{\omega_\kappa(r_{2,3}^2 + 4M^2 - \omega_\kappa^2)}{8r_{2,3}\omega_\kappa(M+K)} \quad (31)$$

$$C_{4,5} = -\frac{1}{2\pi}\frac{1}{e^{\beta r_{4,5}}-1} \times \\ \frac{1}{ip_n + r_{4,5} - \varepsilon_{\mathbf{p}+\kappa}}\frac{\omega_\kappa(r_{4,5}^2 + 4M^2 - \omega_\kappa^2)}{8r_{4,5}\omega_\kappa(M+K)}$$

Where the factors in Eq. (31) are defined as

$$r_1 = -ip_n + \varepsilon_{\mathbf{p}+\kappa} \\ r_2 = \sqrt{(4MK+\omega_\kappa^2)+2(M\omega_\kappa+K\omega_\kappa)} = -r_3 \quad (32) \\ r_4 = \sqrt{(4MK+\omega_\kappa^2)-2(M\omega_\kappa+K\omega_\kappa)} = -r_5$$

$r_5 = -r_4$, respectively.

Eqs. (29)-(32) are the main results of this study, showing that how the electron energy and relaxation time will change when a dislocation starts to interact with another.

## VI. Results and Discussions

We numerically compute Eqs.(29)-(32) to see that how electron self-energy would change with electron-dislocation distance *r* and inter-dislocation distance *d*. We plug in material parameters close to Germanium values as a proof of concept, yet we do not intend to compute realistic materials since the oversimplified electron model and dislocation type. The Poisson ratio we set is $\nu = 0.3$, with the Lame parameters $\lambda = 48$GPa and $\mu = 41$GPa. The mass density and the in-plane Debye wavevector cutoff of the material we set is $\rho = 5.3 g/cm^3$ and $k_D = 5 nm^{-1}$, and the electron energy is taken at $E_0 = 0.67 eV$ which is at the bottom of conduction band minimum, with temperature $T$=10K in Matsubara formulism. The angle between the first dislocation and the electron is $\pi/4$, while the Thomas-Fermi wavevector is $k_{TF} = 3 nm^{-1}$.

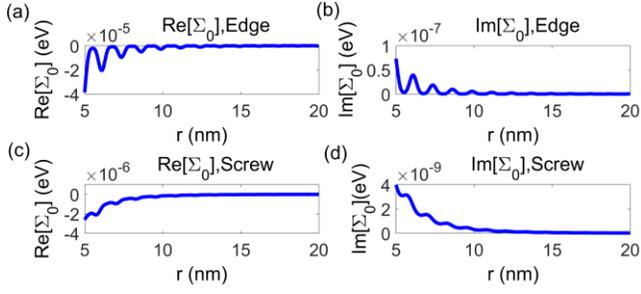

**Fig. 2.** The real and imaginary parts of the electron self-energy when scattering with a single independent dislocation line, for an edge dislocation (a and b) and a screw dislocation (c and d), where r is the distance between the dislocation core and the electron. The oscillation for an edge dislocation is more drastic than that for a screw dislocation, which is caused by a dilatation effect.

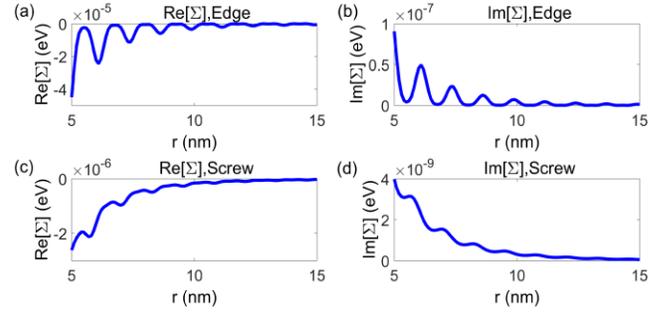

**Fig. 3.** The real and imaginary parts of the electron self-energy when the electron is scattered with interacting dislocations, for a pair of edge dislocations (a and b) and a pair of screw dislocations (c and d). In all figures r is the distance between the dislocation $a$ and the electron. Here the distance between two dislocations is fixed at 10nm, and the angle between the two dislocations is $\phi = \pi/3$.

First, we compute the electron self-energy $\Sigma_0$ with a single, non-interacting dislon as the reference state. Fig. 2 shows that how the real and the imaginary parts of electron self-energy vary as a function of electron-dislocation core distance r, for an edge dislocation (Fig. 2a and 2b) and a screw dislocation (Fig. 2c and 2d), respectively. We observe a more drastic oscillation near an edge dislocation compared with a screw dislocation, which is consistent with the results reported in [11].

The electron self-energy $\Sigma$ which takes into account the inter-dislon interaction as a function of r is plotted in Fig. 3, with fixed inter-dislocation distance d=10nm (corresponding to a ~$10^{12}$/cm² dislocation density) and the angle between the two dislocations fixed at $\phi = \pi/3$ (typical value for partial dislocation pairs), with other parameters taken identical as Fig. 2. The trend of oscillation resembles very much as the non-interacting case in Fig. 2; however, when taking the ratio between the interacting and non-interacting scenarios, the differences are immediately revealed. Since we are more interested in the relaxation rate change (imaginary part) than in the energy shift, the ratio between the electron relaxation rate with the interacting dislons $\text{Im}\Sigma$ and with the non-interacting rate $\text{Im}\Sigma_0$ is plotted in Fig. 4. Surprisingly, it can be seen that when an electron is approaching a dislocation core, its self-energy ratio does not decrease monotonically with the presence of the dislocation-dislocation interaction, but in an oscillatory way. Such an oscillation can be understood as the interference pattern when the localized vibrational modes from individual dislocations begin to superpose upon each other.

Now we could quantify Matthiessen's rule for IDS by changing the dislocation density and thereby the inter-dislocation distance d, as shown in Fig. 5, where we have set the electron-dislocation core distance fixed at r= 7.5nm, which is a value comparable with dislocation-dislocation distances. In addition to the oscillatary behavior, we could see that the electron self-energy asymptotically approaches to the low dislocation density value when d>10nm (the deviation <1%), while when d~5nm, a ~10% deviation for edge dislocations, and ~2% deviation for screw dislocations are seen. Since the d~10nm already corresponds to a highly dislocated crystal with dislocation density $n_{dis} \sim 10^{12}/cm^2$, we conclude that Matthiessen's rule for IDS is still a good approximation even for a highly dislocated crystal. However, with further enhancing the dislocation density, such as dislocation density close to the upper threshold [21], the deviation of Matthiessen's rule will become apparent. This is in qualitatively agreement with the experimental facts that the deviation only occurs in heavily disordered samples [3].

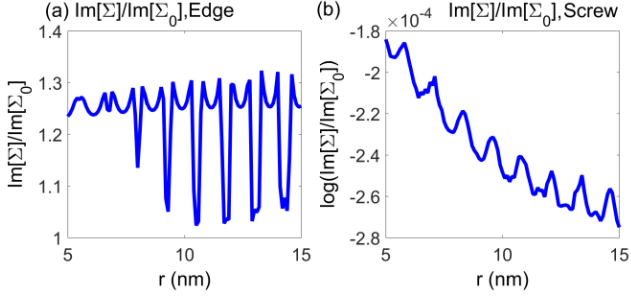

**Fig. 4.** The ratios between the electron relaxation rate with interacting dislons $Im\Sigma$ and the non-interacting rate $Im\Sigma_0$ for edge dislocations(a) and for screw dislocations (b), where r is the distance between the electron and the first dislocation core. The vertical axis of (b) is on a log scale for clarity.

**VII. Conclusions**

In this study, we have presented a quantitative theory to study how interacting dislocations would affect the electron energy and relaxation rate. With the aid of this theory, we could readily address the answers posed by the questions listed at the beginning of this study:

1) When an electron is approaching toward a dislocation core, the deviation from the Matthiessen's rule for IDS increases oscillatively when reducing the electron-dislocation distance. This is caused by the interference between the "ripple" like dislon modes of each dislocation, which is essential to the quantum coherence aspects of the dislons. However, in realistic materials, such an effect might be difficult to observe, given that the dislocation-dislocation distance is not a fixed value but lies in a range of magnitudes, resulting in an overall cancellation of interference pattern and thereby decoherence.

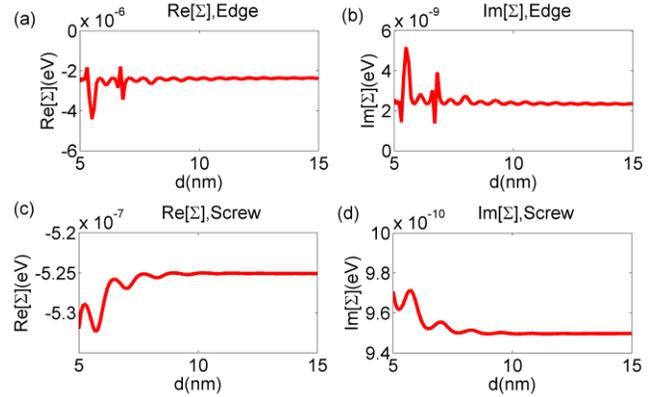

**Fig. 5.** The real and imaginary parts of the electron self-energy as a function of dislocation-dislocation distance d. (a) and (b) are the real and imaginary parts of the self-energy for edge dislocations, while (c) and (d) are the corresponding results for screw dislocations. The distance between the dislocation $a$ and the electron is set fixed at 7.5nm.

2) Similarly, the deviation from Matthiessen's rule for IDS does not grow monotonically as the dislocation density increases, but also varies oscillatively due to the interference effect.

3) Even for a highly-dislocated crystal with a dislocation density as high as $n_{dis} \sim 10^{12}/cm^2$, in the current framework, the use of Matthiessen's rule is for IDS is still a valid approximation with deviation <1%; it is only when the dislocation density becomes even higher than, say hypothetically $n_{dis} \sim 5\times10^{12}/cm^2$, the deviation would become apparent. However, it is worthwhile mentioning that this conclusion is drawn based on only pairwise interaction assumption. In a real crystal, a lower density of deviation may occur when considering strong multiple dislocation-dislocation interactions, i.e. one dislocation interacts with all other dislocations, which drives the system toward a strongly correlated system. In addition, in a highly-disordered crystal, dislocations are not the only type of defects- the interaction between dislocations and other types of defects shall also be taken into account.

To summarize, this fully quantized theory of interacting dislocations could serve as a tool to compute electron transport properties in a highly-dislocated system. We demonstrated the quantitative criteria for whether the dislocation-dislocation interaction is negligible or not in considering electron-dislocation scattering by virtue of dislon, the quasiparticle associated with crystal dislocations at full quantum field theoretical level. Without using a full quantum field theoretical approach as we did in this study, it is unimaginable to study a complex interacting system by merely using a 1$^{st}$-order perturbation approach or a semi-classical approach, which are the traditional approaches in treating electron-dislocation interacting systems.

**Acknowledgements**

ML would thank G. Chen and M. S. Dresselhaus their helpful discussions and the support by S$^3$TEC, an Energy Frontier Research Center funded by U.S. Department of Energy (DOE), Office of Basic Energy Sciences (BES) under Award No. DE-SC0001299/DE-FG02-09ER46577.

To whom correspondence should be addressed:
*mingda@mit.edu